\UseRawInputEncoding
\documentclass[prb,preprint,amsmath,amssymb,superscriptaddress]{revtex4-1}
\usepackage{graphicx}
\usepackage{graphicx}
\usepackage{dcolumn}
\usepackage{bm}
\usepackage{mathrsfs}
\usepackage{subfigure}
\usepackage{natbib}
\usepackage{color}
\usepackage{amssymb}
\usepackage{amsmath}
\usepackage{cancel}
\usepackage{caption}
\usepackage{bbm}
\usepackage{float}
\usepackage[breaklinks,colorlinks,citecolor=blue]{hyperref} 
\hypersetup{%
  colorlinks=true,
  citecolor=blue,
  urlcolor=blue,
  linkcolor=blue,
  }

\usepackage{changes}
    \colorlet{Changes@Color}{blue}

\begin{document}

\title{Robust Superconductivity in Quasi-One-Dimensional Multiband Materials}

\author{T. T. Saraiva}
\affiliation{HSE University, 101000, Moscow, Russia}

\author{L. I. Baturina}
\affiliation{RTU MIREA, Lomonosov Institute of fine chemical technologies, 119454, Moscow, Russia}

\author{A. A. Shanenko}
\affiliation{HSE University, 101000, Moscow, Russia}

\begin{abstract} 
Recently it has been demonstrated that the pair-exchange coupling of quasi-one-dimensional (Q1D) bands with conventional higher-dimensional bands in one multiband superconducting material can result in the formation of robust aggregate pair condensate.  In particular, it has been found that the Q1D thermal pair fluctuations are suppressed in the presence of deep conventional band(s), where the Fermi level is much larger than the characteristic cut-off energy. Here we report that impact of the Q1D fluctuations is significantly weakened even in the presence of nearly shallow higher-dimensional band(s), which shed new light on robust superconducting state observed in emerging chain-like-structured superconducting materials A$_2$Cr$_3$As$_3$ (A = K, Rb, Cs). 
\end{abstract}
\maketitle

Understanding new physics in emerging superconducting functional materials is the key condition for their possible future technological applications. Coexistence of superconductivity with ferromagnetism~\cite{Stolyarov2018, Stolyarov2020}, the formation of vortex clusters and vortex patterns,~\cite{Shanenko2018, Shanenko2021} the interplay between superconductivity and dimensionality as a matter of searching for superconductors with a high transition temperature~\cite{Yan2019, Bianconi2019, Yan2021}, the strain-driven enhancement of superconductivity~\cite{Lu2021} are just few examples of rich related phenomena.

It is well known that superconductivity suffers at lower dimensions due to pronounced fluctuations of the order parameter. However, recently it has been demonstrated that such pair fluctuations are weakened due to interference of multiple condensates that coexist in one multiband superconducting material and have different dimensions.~\cite{Salasnich2019,Saraiva2020}. In particular, it has been demonstrated~\cite{Saraiva2020} that severe Q1D thermal fluctuations are suppressed due to the Cooper-pair transfer between Q1D and 3D bands when the latter is deep enough. In this case the Fermi level of the 3D band is much larger than the characteristic cut-off energy and the pair fluctuations of the corresponding 3D condensate are negligible. In the present work, motivated by these theoretical results and on-going studies of multiband Q1D superconductors A$_2$Cr$_3$As$_3$ (A = K, Rb, Cs)~\cite{Jiang2015,Bao2015,Zhi2015,Tang2015,Wu2019,Xu2020}, we continue investigations of the two-band model with the coupled Q1D and 3D bands. Our present study is focused on the pair fluctuations in the Q1D+3D two-band system beyond the regime of a deep 3D band. 

We consider the two-band generalization of the BCS model, where $s$-wave pair condensates in both bands are coupled via the Cooper-pair transfer~\cite{Suhl1959,Moskalenko1959}. The coupling matrix $g_{\nu\nu'}$ ($\nu,\nu'=1,2$) is symmetric, with $\nu=1$ for the 3D band and $\nu=2$ for the Q1D band. We employ the effective mass approximation so that the single-particle energies (absorbing the chemical potential $\mu$) are given by
\begin{equation}
\xi_{{\bf k}1}=\varepsilon_0+\sum\limits_i\frac{\hbar^2 {\bf k}^2}{2m_1}-\mu, \;\,\xi_{{\bf k}2}=\frac{\hbar^2 k_x^2}{2m_2}-\mu,
\label{eq.xinu}
\end{equation}
where $m_{1,2}$ are the effective carrier masses for bands $1$ and $2$, and ${\bf k}=\{k_x,k_y,k_z\}$. The Q1D energy dispersion is degenerate in the $y$ and $z$ directions and we assume finite contributions to the density of states (DOS) from these directions. All the energies are measured from the lower edge of the Q1D band. We adopt $\varepsilon_0 < 0$ and assume that the standard BCS approximations are applied for band $1$. For illustration, below we consider the domain $|\mu| < \hbar\omega_c$, with $\hbar\omega_c$ the cut-off energy, where the Q1D Feshbach-like resonance determines the system properties. The role of impurities is neglected and we set $k_B=1$ for the Boltzmann constant.

To investigate impact of the Q1D thermal pair fluctuations, we utilize the Gibbs distribution $e^{-F/T}$ with $F$ the free energy. For the model of interest it is given by
\begin{equation}
F = \int d^3{\bf r}\Big[\sum_{\nu=1,2} f_\nu  +  \langle\vec\Delta, \check{L},
\vec\Delta\rangle\Big],   
\label{free}
\end{equation}
where $f_{\nu}$ are the band contributions depending on the gap function $\Delta_{\nu}=\Delta_{\nu}({\bf r})$ and its spatial gradients while the second term controls the Cooper-pair transfer between the bands. In Eq.~(\ref{free}) we use a vector notation $\vec\Delta = (\Delta_1,\Delta_2)$, with $\langle.,.\rangle$ the scalar product in the band vector space, and the matrix $\check{L}$ is defined as
\begin{align}
\check{L}=\check{g}^{-1}
-\left(\begin{array}{cc} {\cal A}_1 & 0\\
0 & {\cal A}_2
\end{array}
\right). 
\label{L}
\end{align}
with $\mathcal{A}_{\nu}$ given by
\begin{align}
\mathcal{A}_{1}=N_1\ln\left(\frac{2e^\Gamma}{\pi\tilde{T}_{c0}}\right), \,
\mathcal{A}_2=N_2\int\limits_{-\tilde\mu}^{1}d\varepsilon\frac{\tanh(\varepsilon/2\tilde{T}_{c0})}{\varepsilon\sqrt{\varepsilon+\tilde\mu}}
\label{A}.
\end{align}
Here the density of states (DOS) for band $1$ is given by $N_1=m_1k_F/2\pi^2\hbar^2$, with the Fermi wavenumber $k_F =\sqrt{2m_1(\mu+|\varepsilon_0|)} /\hbar$, and  for the Q1D band we have $N_2= \sigma_y\sigma_z \sqrt{m_2/32\pi^2 \hbar^3 \omega_c}$ ($\sigma_y$ and $\sigma_z$ account for the contributions from the $y$ and $z$ directions), which is the Q1D DOS at the cut-off energy $\hbar\omega_c$. In addition, $T_{c0}$ is the mean-field critical temperature, the quantities marked by a tilde are given in units of the cut-off energy $\hbar\omega_c$, and $\Gamma\approx 0.577$. 

In the vicinity of $T_{c0}$ the contribution $f_{\nu}$ can be expanded in powers of $\Delta_{\nu}$ and its spatial gradients as  
\begin{equation}
f_\nu = a_\nu \left|\Delta_{\nu} \right|^2  + \frac{b_{\nu}}{2} \left|
\Delta_{\nu} \right|^4 +\sum\limits_{i=x,y,z} {\cal K}^{(i)}_{\nu}  \left|\partial_i\Delta_{\nu}
\right|^2.
\label{fnu}
\end{equation}
For the 3D band ($\nu=1$) the coefficients in Eq.~(\ref{fnu}) are given by the standard expressions
\begin{align}
&a_1= -\tau N_1,\, b_1 = \frac{7\zeta(3)}{8\pi^2}\frac{N_1}{T^2_{c0}},\nonumber\\
&{\cal K}^{(x)}_1={\cal K}^{(y)}_1={\cal K}^{(z)}_1=\frac{\hbar^2v_1^2}{6} b_1,
\label{coeff3DK}
\end{align}
with $\tau=1-T/T_{c0}$ and the band Fermi velocity $v_1=\hbar k_F/m_1$. For the Q1D band ($\nu=2$) we obtain  ($|\mu| < \hbar\omega_c$) 
\begin{align}
a_2=&-\tau\frac{N_2}{2\tilde T_{c0}} \int\limits_{-\tilde{\mu}}^{1}d\varepsilon \, \frac{\text{sech}^2  \big( \varepsilon / 2\tilde T_{c0} \big) }{\sqrt{\varepsilon+\tilde{\mu}}}, \notag \\
b_2 = & \frac{N_2}{4 \hbar^2 \omega_c^2} \int\limits_{-\tilde{\mu}}^{1}d\varepsilon\,  \frac{\text{sech}^2  \big(\varepsilon/2\tilde T_{c0} \big) }{\varepsilon^3 \sqrt{\varepsilon+\tilde{\mu}}}  \left[ \sinh \Big( \frac{\varepsilon}{\tilde T_{c0}} \Big)- \frac{\varepsilon}{\tilde T_{c0}}\right], \notag \\
\mathcal{K}^{(x)}_2 = &\hbar^2 v_2^2  \frac{N_2}{8\, \hbar^2 \omega_c^2} \int\limits_{-\tilde{\mu}}^{1} d\varepsilon\, \frac{\sqrt{\varepsilon+\tilde{\mu}} }{\varepsilon^3} \, \text{sech}^2  \big(\varepsilon / 2\tilde T_{c0} \big)  \notag \\
& \times \left[ \sinh \left( \frac{\varepsilon}{\tilde T_{c0}} \right)- \frac{\varepsilon}{\tilde T_{c0}}\right], \quad  {\cal K}^{(y,z)}_2=0,
\label{coeffQ1D}
\end{align}
where the characteristic velocity $v_2$ for the Q1D band is given by $v_2 = \sqrt{2\hbar\omega_c/m_2}$.

To proceed further, we need to specify the mean-field critical temperature, which is obtained from the linearized gap equation. For the free energy functional (\ref{free}) this equation reads
\begin{align}
\breve{L}
\vec{\Delta}=0.
\label{gap}
\end{align}  
A nontrivial solution for $\Delta$ exists only when the determinant of the matrix $\breve{L}$ is zero so that    
\begin{equation}
\left(g_{22}-G\mathcal{A}_1\right)\left(g_{11}-G\mathcal{A}_2\right)-g_{12}^2=0,
\label{Tc0}
\end{equation}
where $G=g_{11}g_{22}-g_{12}^2$. As ${\cal A}_\nu \propto N_\nu$, one concludes that the couplings $g_{\nu\nu'}$ and the DOSs $N_{\nu}$ can be combined in a smaller set of parameters so that $T_{c0}$ depends only on the three dimensionless couplings
\begin{equation}
\lambda_{11}=g_{11}N_1,\;
\lambda_{22}=g_{22}N_2,\;
\lambda_{12}=g_{12}\sqrt{N_1N_2}.
\label{lambda}
\end{equation}
Equation~(\ref{Tc0}) has two solutions for $T_{c0}$, and the highest one should be chosen among them as it corresponds to a lower free energy of the system. 

One can significantly simplify the expression for the free energy (\ref{free}) by representing $\vec{\Delta}$ as a linear combination of the eigenvectors of the matrix $\breve{L}$ in the form~\cite{Salasnich2019,Saraiva2020}
\begin{equation}
\vec{\eta}_+=
\left(\begin{array}{c}
S\\
1
\end{array}\right),\;\vec{\eta}_-=
\left(\begin{array}{c}
1\\
-S
\end{array}\right),
\label{eta}
\end{equation}
with $S$~($S \geq 0$ for the $s$-wave pairing) defined as
\begin{align}
S=\frac{g_{11}-G\mathcal{A}_2}{g_{12}}.
\label{S}
\end{align}
Then we have
\begin{align}
\vec{\Delta}({\bf r}) = \psi ({\bf r})\vec{\eta}_+ +\varphi({\bf r}) \vec{\eta}_-,
\label{Dexp}
\end{align}
where $\psi({\bf r})$ and $\varphi({\bf r})$ are the condensate modes associated with $\vec{\eta}_+$ and $\vec{\eta}_-$. One can check~\cite{Salasnich2019,Saraiva2020} that only $\psi$ is the critical mode with the divergent characteristic length and fluctuation het capacity at $T \to T_{c0}$. The fluctuations controlled by the mode $\varphi$ produce noncritical corrections and can be safely neglected. Thus, removing all the terms including $\varphi$, we arrive at the single-component Ginzburg-Landau functional 
\begin{equation}
F =\int d^3{\bf r}\Big(a_\psi |\psi|^2  + \frac{b_{\psi}}{2}|\psi|^4 +\sum\limits_{i=x,y,z} {\cal K}^{(i)}_{\psi}|\partial_i\psi|^2\Big),  
\label{freepsi}
\end{equation}
with 
\begin{align}
&a_{\psi}=S^2  a_1 + a_2,\; b_{\psi} = S^4 b_1 + b_2,\notag\\
&{\cal K}^{(i)}_{\psi} = S^2{\cal K}^{(i)}_1 + {\cal K}^{(i)}_2,
\label{coeffpsi}
\end{align}
where the presence of the two bands is reflected in that the coefficients $a_\psi, b_\psi$, and ${\cal K}^{(i)}_{\psi}$ are the averages over the available bands, see Eq.~(\ref{coeffpsi}). 
 
Impact of the thermal pair fluctuations is measured by the Ginzburg number $Gi$ (also the Ginzburg-Levanyuk parameter). We recall that $Gi=1-T_{Gi}/T_{c0}$, where $T_{Gi}$ is the temperature at which the mean-field heat capacity is equal to the fluctuation-driven heat capacity~\cite{Larkin} and at $T > T_{Gi}$ the fluctuations contribution predominates.  The Ginzburg number for the free energy functional (\ref{freepsi}) is expressed as~\cite{Larkin,Saraiva2020} 
\begin{equation}
Gi=\frac{1}{32\pi^2}\frac{T_{c0}b_\psi^2}{a^\prime_\psi
{\cal K}^{(x)}_\psi {\cal K}^{(y)}_\psi {\cal K}^{(z)}_\psi}.
\label{Gi12}
\end{equation}
where $a^\prime_\psi=da_\psi/dT$. Utilizing Eq.~(\ref{coeffpsi}), we obtain
\begin{equation}
Gi=Gi_1 \frac{(b_2/b_1+S^4)^2}{S^4 (a'_2/a'_1+S^2) \big(\mathcal{K}^{(x)}_2/\mathcal{K}^{(x)}_1+S^2\big)},
\label{Gi12A}
\end{equation}
where $a^\prime_\nu=da_\nu/dT$ and 
\begin{equation}
Gi_1 =\frac{1}{32\pi^2}\frac{T_{c0}b_1^2}{a^\prime_1
\mathcal{K}^{(x)}_1 \mathcal{K}^{(y)}_1 \mathcal{K}^{(z)}_1}= \frac{27\pi^4}{14\zeta(3)}\left(\frac{\tilde{T}_{c0}}{\tilde\mu+|\tilde\varepsilon_0|}\right)^4
\label{Gi1}
\end{equation}
is the Ginzburg number of band $1$. 

It is of importance to examine the limiting values of the Ginzburg number, using Eqs.~(\ref{Gi12A}). As is mentioned above, when $S \to \infty$, band $2$ does not contribute [see Eq.~(\ref{eta})] so that the superconducting properties are determined by band $1$. In this case Eq.~(\ref{Gi12A}) is naturally reduced to $Gi=Gi_1$. For $S \to \infty$ the contribution of band $1$ becomes negligible (the unstable Q1D condensate predominates) and $Gi \to \infty$. This is not consistent with the definition of the Ginzburg number, according to which $Gi \leq 1$. In fact, the origin of this deviation is not surprising since in the Q1D case we deal with the situation when two stiffness GL coefficients ${\cal K}^{(y)}_1$ and ${\cal K}^{(z)}_2$ go to zero. Indeed, for $S \to \infty$ one gets ${\cal K}^{(y)}_\psi = S^2{\cal K}^{(y)}_1 \to 0$ and ${\cal K}^{(z)}_\psi = S^2 {\cal K}^{(z)}_1 \to 0$. It means that for the two spatial directions the integral over the momentum in the fluctuation-driven heat capacity becomes divergent, see details of the calculations of the fluctuation contribution to the heat capacity in Ref.~\onlinecite{Larkin}. To avoid such an artificial divergence, the momentum cut-off should be introduced in the integration (i.e., the boundary of the Brillouin zone). Since we are interested in the regime when $Gi$ is significantly smaller than $1$~(the pair fluctuations are not pronounced), we can simply ignore this cut-off and related technical complications. It is enough to keep in mind that Eqs.~(\ref{Gi12}) and (\ref{Gi12A}) are not applicable when the Ginzburg number approaches $1$.     

Though the Ginzburg number is an important характеристика of the pair thermal fluctuations, it is not an observable. Then,  it is more convenient to employ the fluctuation-shifted critical temperature $T_c$ that is a function of $Gi$. We can utilize the 3D renormalization group result in the form~\cite{Larkin}
\begin{equation}
\frac{T_c-T_{c0}}{T_c}=\frac{8}{\pi}\sqrt{Gi},
\label{Tc}
\end{equation}
which is used below together with Eq.~(\ref{Gi12A}). 

\addtocounter{figure}{-1}
\begin{figure*}
	\begin{center}
		\includegraphics[width=0.7\textwidth]{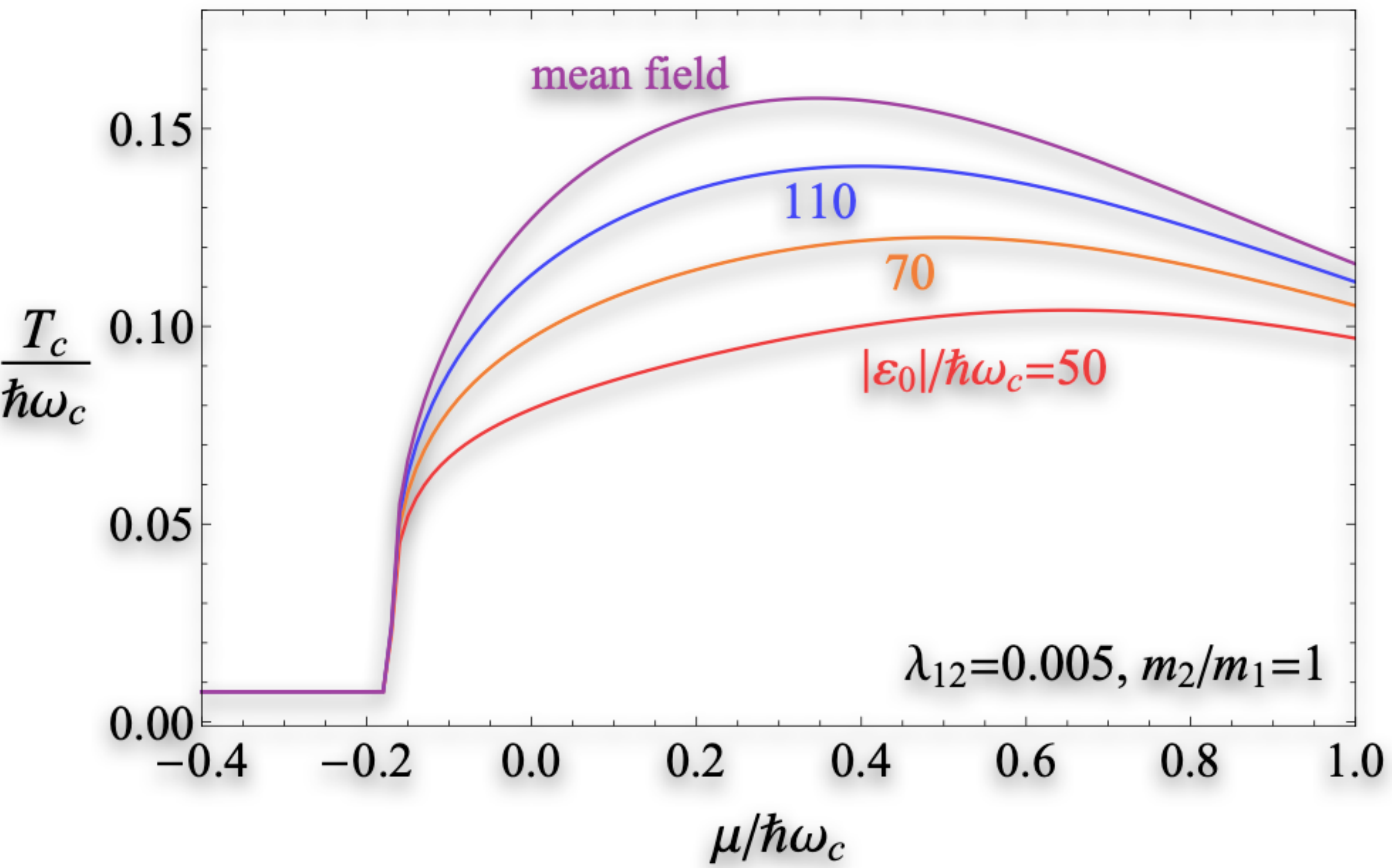}
	\end{center}
	\caption{The mean-field $T_{c0}$ and fluctuation-shifted $T_c$ critical temperatures 
	as functions of the chemical potential $\mu$ calculated for $|\varepsilon_0|/\hbar\omega_c=50,70$, and $110$ 
	at $\lambda_{12}=0.005$ and $m_2/m_1=1$.  Other microscopic parameters are given in the text. }
	\label{fig1}
\end{figure*}

As is discussed above, the relative mean-field critical temperature $\tilde{T}_{c0}$~(in units of $\hbar\omega_c$) is determined by the three dimensionless coupling constants $\lambda_{11},\lambda_{22}$, $\lambda_{12}$ and the chemical potential $\tilde{\mu}$. The fluctuation-shifted critical temperature $\tilde{T}_c$ involves additional microscopic parameters. From Eq.~(\ref{Gi12A}) we find that the Ginzburg number of the two-band system depends on $Gi_1$, $S$ and the ratios $N_2/N_1$ and $v_2/v_1$. As is seen from Eq.~(\ref{S}), $S$ is determined by $\tilde{T}_{c0}$ and $N_2/N_1$ and, keeping in mind our results for $\tilde{T}_{c0}$, we find that $S$ depends on $\lambda_{\nu\nu'}$, $\tilde{\mu}$, and $N_2/N_1$. In turn, $Gi_1$ is determined by $\tilde{T}_{c0}$, $\tilde{\mu}$ and $\tilde{\varepsilon}_0$, see Eq.~(\ref{Gi1}). Finally, the ratio $v_2/v_1=\sqrt{m_1/\big(m_2(\tilde{\mu} + |\tilde{\varepsilon}_0|)\big)}$ is controlled by $\tilde{\mu}$, $\tilde{\varepsilon}_0$, and the ratio of the band masses $m_2/m_1$. Thus, Eqs.~(\ref{Tc}) and (\ref{Gi12A}) prescribe that $\tilde{T}_c$ is determined by the following set of the dimensionless parameters: $\lambda_{\nu\nu'}$, $\tilde{\mu}$, $\tilde{\varepsilon}_0$, $N_2/N_1$, and $m_2/m_1$. 

For illustration we choose $\lambda_{22}=0.18$ and $\lambda_{11}=0.2$ which correspond to dimensionless couplings of conventional $s$-wave superconductors~\cite{Fetter}. Usually in two-band systems we have~\cite{Vagov2016} $\lambda_{12} \ll \lambda_{11},\lambda_{22}$. Here we consider $\lambda_{12}=0.005,0.01$ and $0.02$. The ratio of the band DOSs $N_2/N_1$ is close to $1$ in most cases~\cite{Salasnich2019}, and below we set $N_2/N_1=1$. Finally, $\tilde{\mu}$, $\tilde{\varepsilon}$, and $m_2/m_1$ can change significantly from one material to another, depending on fabrication details and doping, and we treat these quantities as free parameters.   

Figure~\ref{fig1} presents $\tilde{T}_{c0}$ and $\tilde{T}_c$ versus $\tilde{\mu}$, as calculated for $|\tilde{\varepsilon}_0|=50,70$, and $110$ at $\lambda_{12}=0.005$ and $m_2/m_1=1$. One can see that the mean-field critical temperature exhibits a significant increase when the chemical potential approaches the lower edge of the Q1D band and the Q1D Feshbach-like resonance develops.  However, this increase can be compromised by the pair fluctuations.  In particular,  our results demonstrate that the fluctuation-shifted critical temperature approaches the mean-field one only when band $1$ becomes deep enough, which is in agreement with the previous results~\cite{Saraiva2020}. The reasons for this are as follows. First, $Gi_1$ decreases with increasing $\tilde{\varepsilon}_0$, as is seen from Eq.~(\ref{Gi1}). Second, the ratio $v_2/v_1$ becomes larger and so does the ratio ${\cal K}^{(x)}_2/{\cal K}^{(x)}_1$. The latter appears in the denominator of Eq.~(\ref{Gi12A}) so that an increase of ${\cal K}^{(x)}_2/{\cal K}^{(x)}_1$ leads to a decrease of $Gi$. As a result, the pair fluctuations are eventually suppressed and $\tilde{T}_c$ approaches $\tilde{T}_{c0}$, as follows from Eq.~(\ref{Tc}). 

\begin{figure}
	\begin{center}
		\includegraphics[width=0.7\textwidth]{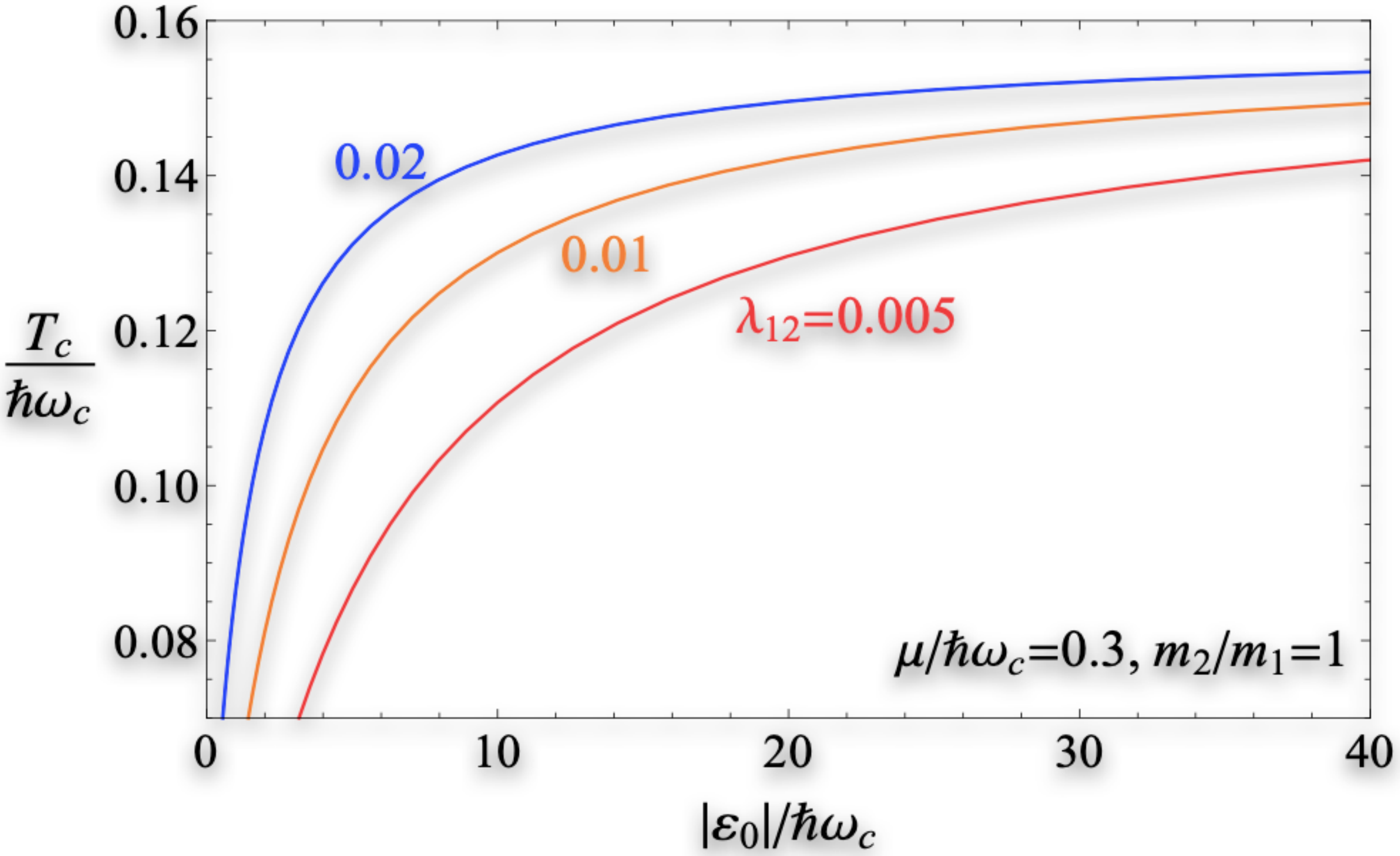}
	\end{center}
	\caption{The mean-field $T_{c0}$ and fluctuation-shifted $T_c$ critical temperatures as functions of $|\varepsilon_0|$ calculated for $\mu=0.3\hbar\omega_c$ and $\lambda_{12}=0.005, 0.01, 0.02$ at $m_2/m_1=1$. Other microscopic parameters are discussed in the text.}
\label{fig2}
\end{figure}  

In the opposite regime of decreasing $|\tilde{\varepsilon}_0|$, the 3D condensate tends to lose its robustness and then, at sufficiently small $|\tilde{\varepsilon}_0|$, it cannot stabilize the aggregate condensate of the two-band system compromised by the Q1D pair fluctuations. The resulting critical temperature drops, deviating  significantly from the mean-field one, and we arrive at the regime of strong pair fluctuations suppressing the superconducting correlations. This raises the question of how small $|\tilde{\varepsilon}_0|$ can be to still have moderate thermal fluctuations in the two-band system. 

\begin{figure}
\begin{center}
\includegraphics[width=0.7\textwidth]{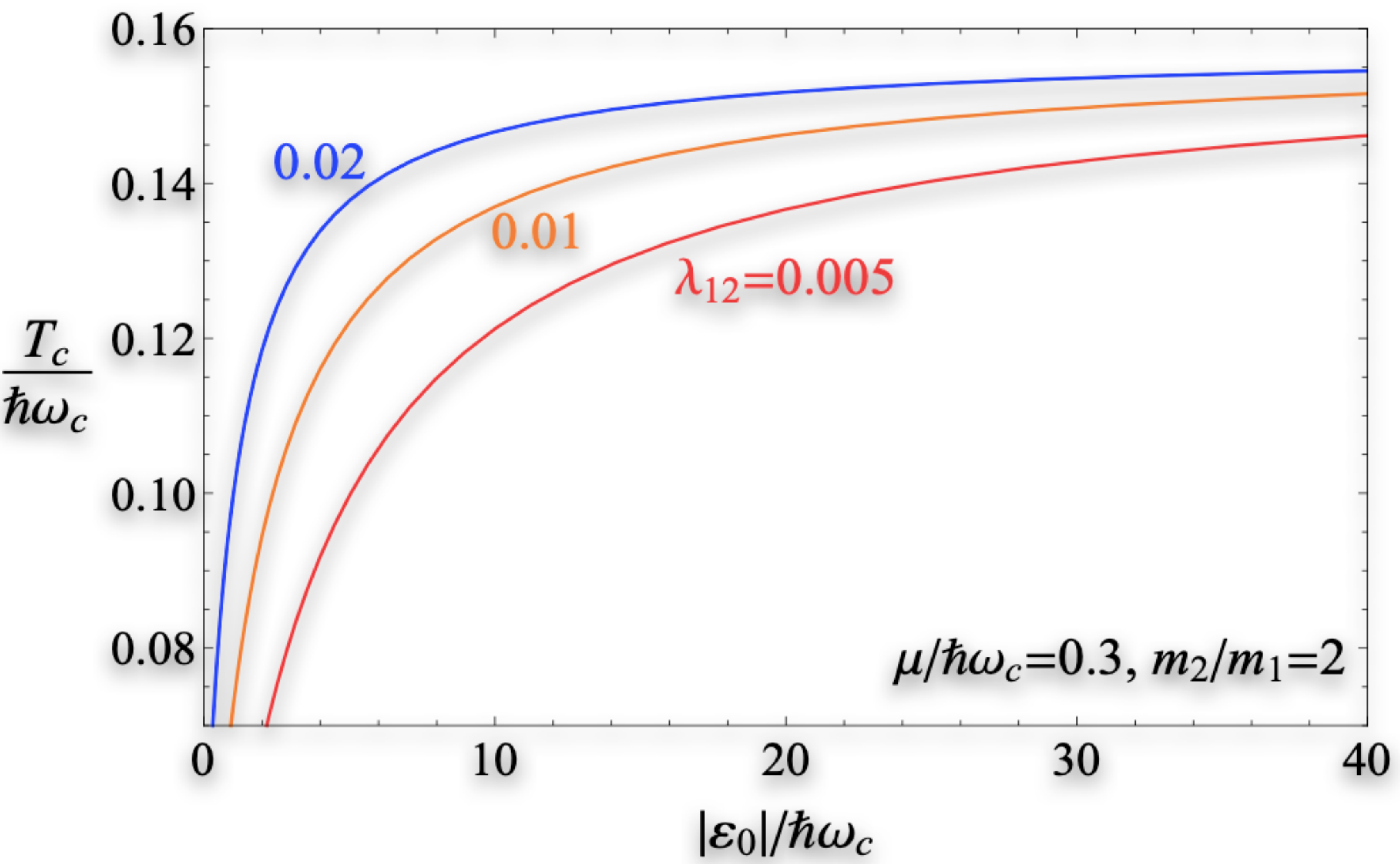}
\end{center}
 	\caption{The same as in Fig.~\ref{fig2} but now for $m_2/m_1=2$.}
\label{fig3}
\end{figure} 

To answer this question, we investigate $\tilde{T}_c$ as a function of $|\tilde{\varepsilon}_0|$ at the fixed chemical potential $\tilde{\mu}=0.3$~(near the maximum of $\tilde{T}_{c0}$). In particular, Fig.~\ref{fig2} presents $\tilde{T}_c$ versus $|\tilde{\varepsilon}_0|$ for the three different values of the pair-exchange coupling $\lambda_{12}=0.005, 0.01$, and $0.02$ at $m_2/m_1=1$.  In agreement with Fig.~\ref{fig1},  $\tilde{T}_c$ increases with $|\tilde{\varepsilon}_0|$ and at sufficiently large values of  $|\tilde{\varepsilon}_0|$ the fluctuation-shifted critical temperature approaches the mean-field superconducting temperature  $\tilde{T}_{c0}\approx 0.155$. The latter is almost independent of $\lambda_{12}$ as the mean-field temperature is mainly governed by the Q1D Feshbach-like resonance~\cite{Saraiva2020}.  At small $|\tilde{\varepsilon}_0|$ we find that $\tilde{T}_c$ deviates significantly from $\tilde{T}_{c0}$. However, surprisingly, a sharp drop of $\tilde{T}_c$ occurs only in the region $|\tilde{\varepsilon}_0| \lesssim 10$. In particular,  one finds that the fluctuation-shifted critical temperature can still be significant~(e.g., larger than a half of $\tilde{T}_{c0}$) for a nearly shallow band $1$, where $|\tilde{\varepsilon}_0| \approx 0.5$-$1.0$, as is seen for $\lambda_{12}=0.02$. Notice that in this case the relative Fermi energy $\tilde{E}_F=|\tilde{\varepsilon}_0|+\tilde{\mu}$~(in units of $\hbar\omega_c$) is about $1$-$2$ and thus, we arrive at the estimate $E_F/T_c \approx 10$-$20$, which is orders of magnitude smaller than the corresponding values in conventional superconductors.

This surprising feature of our results becomes even more pronounced when we adopt larger values of $m_2/m_1$. In particular, Fig.~\ref{fig3} demonstrates the same as Fig.~\ref{fig2} but now for $m_2/m_1=2$. Here we find for $\lambda_{12}=0.02$ that $\tilde{T}_c$ is larger than a half of $\tilde{T}_{c0}$ for $|\tilde{\varepsilon}_0| > 0.3$. For this case $E_F/T_c \approx 7$, which is very close to the BCS-BEC crossover regime in band $1$. Notice that in this case the critical temperature enhancement due to the Q1D Feshbach-like resonance is still by an order of magnitude. Indeed, $\tilde{T}_c$ is about $0.007$ when $\tilde{\mu}$ is below $-0.2$.

Concluding, our results for the two-band model with Q1D and 3D bands demonstrate that severe thermal fluctuations at the Q1D Feshbach-like resonance are notably weakened even in the presence of a nearly shallow 3D band with the band Fermi level of about $7$-$10 T_c$. This shed new light on robust superconducting state observed in recent chain-like-structured superconducting materials A$_2$Cr$_3$As$_3$ (A = K, Rb, Cs) and can encourage further experiments aimed at reaching larger critical temperatures in such materials due to utilizing the Q1D Feshbach-like resonance.  
 
{\it Acknowledgements.} The work at HSE University (A.A.S.) was supported by the project 21-04-041 “Self-organized structures in microscopic quantum systems” in the framework of Program “Scientific Funding of the National Research University “Higher School of Economics” (HSE)” in 2021.

\end{document}